\documentclass[twocolumn,english,prd,tightenlines,floatfix,showpacs,preprintnumbers,nofootinbib,eqsecnum,superscriptaddress]{revtex4-1}
\usepackage[utf8]{inputenc}
\usepackage{color}
\usepackage{babel}
\usepackage{amsmath}
\usepackage{amssymb}
\usepackage{subfig}
\usepackage{graphicx}
\usepackage{paralist}
\usepackage{xspace}
\usepackage[unicode=true,
 bookmarks=false,
 breaklinks=false,pdfborder={0 0 1},backref=section,colorlinks=false]
 {hyperref}

\newcommand{\pyrate}{PyR@TE\xspace}

\newcommand{\SU}[1]{\ensuremath{\mathrm{SU}(#1)}}
\renewcommand{\U}[1]{\ensuremath{\mathrm{U}(#1)}}

\newcommand{\rep}[1]{\ensuremath{\boldsymbol{#1}}}

\def\CC{{C\nolinebreak[4]\hspace{-.05em}\raisebox{.4ex}{\tiny\bf ++}}\xspace}
\DeclareCaptionFont{white}{\color{white}}
\DeclareCaptionFormat{listing}{%
  \parbox{\textwidth}{\colorbox{gray}{\parbox{\textwidth}{#1#2#3}}\vskip-4pt}}

\makeatletter
\def\@fnsymbol#1{%
 \ensuremath{%
  \ifcase#1\or 
   *\or
   \dagger\or
   \ddagger\or
   \mathsection\or
   \mathparagraph\or
   **\or
   \dagger\dagger\or
   \ddagger\ddagger\or %%% CHANGED TO OR 
  \mathsection\mathsection\or
   \mathparagraph\mathparagraph\or
   ***\or
   \dagger\dagger\dagger\or
   \ddagger\ddagger\ddagger\or %%% CHANGED TO OR 
   \mathsection\mathsection\mathsection\or
   \mathparagraph\mathparagraph\mathparagraph\or
   \@ctrerr 
  \fi
 }%
}%
\makeatother
\begin{document}
%%%%%%%%%%%%%%%%%%%%%%%%%%%%%%%%%%%%%%%%%%%%%%%%%%%%%%
%%%%%%%%%%%%%%%%%%%%%%%%%%%%%%%%%%%%%%%%%%%%%%%%%%%%%%

\preprint{SMU-HEP-16-10}

\title{On the impact of kinetic mixing in beta functions at two-loop}

\author{F.~Lyonnet}
\email{flyonnet@smu.edu}
\affiliation{Southern Methodist University, Dallas, TX 75275, USA }

\date{\today}

\begin{abstract}
	Kinetic mixing is a fundamental property of models with a gauge symmetry involving several \U{1} group factors.	In this paper, we perform a numerical study of the impact of kinetic mixing on beta functions at two-loop. To do so, we use the recently published PyR@TE 2 software to derive the complete set of RGEs of the SM B-L model at two-loop including kinetic mixing. We show that it is important to properly account for kinetic mixing as the evolution of the parameters with the energy scale can change drastically. In some cases, these modifications can even lead to a different conclusion regarding the stability of the scalar potential.
\end{abstract}


\maketitle
\tableofcontents{}
\makeatletter
\let\toc@pre\relax
\let\toc@post\relax
\makeatother

\section{Intro}
Renormalization group equations (RGEs) play a central role in studying the high-energy behaviour of extensions of the Standard Model (SM). The RGEs for an arbitrary gauge field theory at two-loop have been known for a long time~\cite{Machacek:1983tz,Machacek:1983fi,Machacek:1984zw,Jack:1982hf,Jack:1982sr,Jack:1984vj} and their derivation for specific models has been automated in~\cite{Lyonnet:2013dna, Staub:2013tta}. An extra complication arises when the gauge structure contains multiple \U{1} group factors as kinetic mixing can occur. This introduces $\frac{1}{2}n(n-1)$ extra dynamical parameters for each of which RGEs must be calculated. In addition, kinetic mixing induces modifications to the RGEs of the other parameters that also need to be taken into account for consistency. This work was carried out in~\cite{delAguila:1988jz,delAguila:1987st} for the gauge couplings and extended in~\cite{Luo:2002iq} for the dimensionless parameters at two-loop. Recently, an alternative method was presented in~\cite{Fonseca:2013jra} in which the modifications to the RGEs for the dimensionful parameters are also derived. We implemented the method of~\cite{Fonseca:2013jra} in the new version~\cite{Lyonnet:2016xiz} of the computer code \pyrate~\cite{Lyonnet:2013dna} which allows us to simply derive RGEs for a given model at two-loop taking into account kinetic mixing.

In this note we show that the impact of such kinetic mixing can be important for the high-energy behaviour of the theory parameters. To do so, we study the RGEs of the SM B-L model in which the SM has been supplemented by an additional $\U{1}_{B-L}$ group factor. We start in Sec.~\ref{sec:kinmix} by giving some details on how the kinetic mixing is implemented in \pyrate. Sec.~\ref{sec:blex} summarizes the principle properties of the SM B-L model while Sec.~\ref{sec:running} is devoted to the study of the impact of kinetic mixing on the running of its parameters. Finally, our conclusions are presented in Sec.~\ref{sec:conclusion}.

\section{Kinetic Mixing in \pyrate\label{sec:kinmix}}

Let's first consider a gauge field theory with $n$ \U{1} group factors $\mathcal{G}\times\U{1}_1\times\U{1}_2\times\dots\times\U{1}_n$ where $\mathcal{G}$ corresponds to the non Abelian part of the gauge structure. Gauge invariance then allows one to write the following term in the Lagrangian:
\begin{equation}
	\mathcal{L}\supset \vec{F}^{\mu\nu}\xi \vec{F}_{\mu\nu}\, ,
\end{equation}
where $\xi$ is an $n\times n$ symmetric matrix, and $F^{\mu\nu}$ is the vector of the $n$ field strength tensors associated to the $n$ gauge groups $\U{1}_i,\ i=1\dots n$. This indeed introduces $\frac{1}{2}n(n-1)$ extra dynamical parameters. These additional parameters can be exchanged by corresponding \textit{effective gauge couplings} that populate the off-diagonal entries of an extended gauge coupling matrix $G=\tilde{G}\xi^{-1/2}$~\cite{Fonseca:2013jra}. Of course the two approaches are equivalent and one can recover the results obtained working with $\xi$ by performing suitable rotations.

The advantage of the effective gauge coupling method is that it does not require the introduction of new parameters $\xi$ and their RGEs but only promotes the gauge couplings to a non-diagonal matrix. The modifications due to kinetic mixing on the beta functions of the other parameters can be expressed as more or less complex replacement rules~\cite{Fonseca:2013jra}. For instance one has 
\begin{equation}
	g^{3}S(R)\rightarrow G\sum_p W_p^R (W_{p}^R)^T\, ,
\end{equation}
with $S(R)$ the Dynkin index of representation $R$. The sum $p$ runs over all the fermions ($R=F$) or scalars ($R=S$). Finally, $W_{i}^R\equiv G^T Q_i^R$, with $Q_i^R$ the vector of Abelian charges of the field $i$ in the representation $R$. 
All the replacement rules derived in~\cite{Fonseca:2013jra} have been implemented in the computer code \pyrate at two-loop.

\section{SM B-L\label{sec:blex}}

Our goal is to assess the size of the modifications due to kinetic mixing on the running of Lagrangian parameters. In order to do so, we will concentrate on the SM extended by an additional $\U{1}_{B-L}$ group factor, i.e. \hbox{B-L} is promoted to a gauge symmetry. We will assume that we have an extra singlet scalar $\chi$ transforming as $~\sim(1,1,0,2)$ under $\SU{3}_c\times\SU{2}_L\times\U{1}_Y\times\U{1}_{B-L}$. In addition, we will consider three right-handed neutrinos $\sim(1,1,0,-1)$. This model has been extensively studied, in particular in the context of dark matter modeling, and constraints on the parameter space have been recently derived in~\cite{Klasen:2016qux}.

\subsection{Lagrangian and symmetry breaking} 

Table~\ref{tab:particlecontent} list all the particles in the model along with their quantum numbers under $\SU{3}_c\times\SU{2}_L\times\U{1}_Y\times\U{1}_{B-L}$ gauge symmetry. Note that our setup is identical to~\cite{Coriano:2015sea}.

\begin{table}
	\centering
	\begin{tabular}{cl}
		Field & Quantum Numbers\\\hline
		$Q_L$ & (\rep{3},\rep{2},\rep{1/6},\rep{1/3})\\
		$u_R$ & (\rep{3}, \rep{1}, \rep{2/3}, \rep{1/3})\\
		$d_R$ & (\rep{3}, \rep{1}, \rep{-1/3}, \rep{1/3})\\
		$L_L$ & (\rep{1}, \rep{2}, \rep{-1/2}, \rep{-1})\\
		$e_R$ & (\rep{1}, \rep{1}, \rep{-1}, \rep{-1})\\
		$\nu_R$ & (\rep{1}, \rep{1}, \rep{0}, \rep{-1})\\[0.1cm]
		$H$ & (\rep{1},\rep{2}, \rep{1/2}, \rep{0})\\
		$\chi$ & (\rep{1}, \rep{1}, \rep{0}, \rep{2})
	\end{tabular}
	\caption{Particle content of the SM B-L model and their quantum numbers under $\SU{3}_c\times\SU{2}_L\times\U{1}_Y\times\U{1}_{B-L}$.}
	\label{tab:particlecontent}
\end{table}
%%%%% 
% Z' mass -> v' 
% lambda 1 lambda 2 lambda 3 as functions of mh1 and mh2 the eigenvalues (inputs: mh1=125 and mh2 fix)
%  
%
With two scalars the most general scalar potential contains 5 parameters and reads
\begin{eqnarray}
	V(H,\chi) &=& \mu_H	H^{\dagger}H + \mu_{\chi}\chi^{\dagger}\chi + \lambda_{1} \left( H^{\dagger}H \right)^{2}\nonumber\\
	&+& \lambda_{2}\left( \chi^{\dagger}\chi \right)^{2}+\lambda_3 \left( H^{\dagger} H \right)\left( \chi^{\dagger}\chi \right)\, .
	\label{eq:potential}
\end{eqnarray}

The two scalars take the following vacuum expectation values (VEVs), leading to spontaneous symmetry breaking (SSB)
\begin{eqnarray}
	<H> = \frac{1}{\sqrt{2}} \left(\begin{matrix} 0\\ v\end{matrix} \right)\, ,<\chi> = \frac{v^{\prime}}{\sqrt{2}}\, .
\end{eqnarray}
The minimization conditions lead to the following expressions for $v$ and $v'$
\begin{eqnarray}
	v^{2}&=&  \frac{\mu_{\chi}^{2}\lambda_3/2 - \mu_{H}^{2}\lambda_2}{\lambda_1\lambda_2- \lambda_3^2/4}\, ,\label{eq:v}\\
	{v'}^2 &=& \frac{\mu_{H}^{2}\lambda_3/2 - \mu_{\chi}^{2}\lambda_1}{\lambda_1\lambda_2- \lambda_3^2/4}\, .\label{eq:vp}
\end{eqnarray}

After SSB, the two neutral scalars mix leading to two physical states of mass $m_{1,2}$. Defining the mixing angle between the two scalars, $\theta$, one can derive the following relations~\cite{Coriano:2015sea}

\begin{eqnarray}
	\lambda_{1}&=&\frac{m_1^{2}}{4v^2}\left( 1+\cos 2\theta \right) + \frac{m_2^{2}}{4v^{2}}\left( 1-\cos 2\theta \right)\, ,\nonumber\\
	\lambda_{2}&=&\frac{m_1^{2}}{4v^{\prime 2}}\left( 1-\cos 2\theta \right) + \frac{m_2^{2}}{4v^{\prime 2}}\left( 1+\cos 2\theta \right)\, ,\nonumber\\
	\lambda_{3}&=&\sin2\theta\left(\frac{m_2^2-m_1^2}{2v v^\prime}\right)\, .
	\label{eq:lbds}
\end{eqnarray}

There is an additional mixing angle, $\alpha$, in the gauge sector which has been tightly constrained by LEP~\cite{Abreu:1994ria}, $|\alpha|\leq 10^{-3}$, and the masses of $Z$ and $Z'$ bosons in this limit can be approximated by 
\begin{equation}
	M_{Z}\simeq	\frac{v}{2}\sqrt{g_2^2 + g^2}\, ,M_{Z'}\simeq \frac{v}{2}\sqrt{\tilde{g}^2 + (4g'v'/v)^2}\, ,
	\label{eq:mzp}
\end{equation}
in which $g_2$ is the gauge coupling of the $\SU{2}_L$ gauge group and $\left(\begin{matrix}g&\tilde{g}\\0&g'\end{matrix}\right)$ are the Abelian gauge couplings in the upper triangular basis which is linked to $G\equiv \left(\begin{matrix}g_{11}& g_{12}\\g_{21} & g_{22}\end{matrix}\right)$ via the rotation
\begin{eqnarray}
	\tilde{G} &=& G\cdot \left(\begin{matrix} \cos \phi & -\sin \phi\\ \sin\phi & \cos \phi \end{matrix}\right),\ \label{eq:ggtilde}\\
\cos\phi &=& \frac{g_{22}}{\sqrt{g_{22}^{2} + g_{21}^2}}\, ,\ \sin\phi \frac{-g_{21}}{\sqrt{g_{22}^{2} + g_{21}^{2}}}\, .\label{eq:cosphi}
\end{eqnarray}

Finally, the Yukawa interactions of the model are dictated by the following Lagrangian

\begin{eqnarray}
	-\mathcal{L}_Y &=&  Y^{ij}_d \overline{Q_L^i}H d^{j}_R + Y_{u}^{ij}\overline{Q^i_L}\tilde{H}u^{j}_R  + Y_{e}^{ij}\overline{L^i} H e^j_R \nonumber\\
	&+& Y_{\nu}^{ij} \overline{L^i}\tilde{H}\nu_R^j + Y_N^{ij} \overline{(\nu^i_R)^c}\nu_R^j\chi + \mathrm{h.c.}\, .
	\label{eq:yuk}
\end{eqnarray}
After SSB this leads to the following mass term for the neutrinos
\begin{equation}
	-\mathcal{L}^{\nu}_{Y} = \underbrace{Y^{ij}_{\nu}\frac{v}{\sqrt{2}}}_{M_d^{ij}}(\nu^{i}_L)^{c} \nu_R^j + \underbrace{\frac{1}{\sqrt{2}}Y^{ij}_{N}v'}_{\frac{1}{2}M_m^{ij}}\nu_R^i\nu_R^j + \mathrm{h.c.}\, ,\label{eq:lagmnu}
\end{equation}
which requires $Y_{\nu}\sim O(10^{-6})$ for light neutrinos while $Y_{N} \sim \mathcal{O}(1)$ for heavy right-handed neutrinos in the TeV range, see~\cite{Coriano:2015sea}. 

\subsection{Parameters and stability}

In the numerical analysis, we will neglect the Yukawa couplings, $Y_e$, $Y_d$, $Y_\nu$, and retain only the top Yukawa coupling,\footnote{Where $y_t$ is third diagonal entry of $Y_u^{33}$.} $y_t$. For simplicity, the right-handed neutrino Yukawa will be reduced to $Y_{N}^{ij}=\delta^{ij}y_N$.

For the numerical analysis, we select the following set of parameters
\begin{equation}
	\mathcal{B}=\{\theta,M_{Z'},M_m, g', \tilde{g}, m_1, m_2\}\, ,\label{eq:benchdef}
\end{equation}
from which we derive initial values for $\{v',y_N, \lambda_1,\lambda_2,\lambda_3,\mu_{H},\mu_{\chi}\}$. Indeed, $v'$ can be extracted from Eq.~(\ref{eq:mzp}) and the knowledge of $M_{Z'},\ g'$ and $\tilde{g}$; $y_N$ is obtained directly from Eq.~(\ref{eq:lagmnu}) whereas the values of the quartic couplings $\lambda_1,\lambda_2,\lambda_3$ result from Eq.~(\ref{eq:lbds}). Finally, the initial values for the Lagrangian mass parameter, $\mu_{H}$ and $\mu_\chi$ are fixed via the minimization conditions, Eq.(\ref{eq:v}) and Eq.(\ref{eq:vp}). The SM Higgs mass is identified with the physical scalar mass $m_1=126$ GeV. 

It is well known, that the stability of the scalar potential is achieved by the following conditions~\cite{Coriano:2015sea}
\begin{equation}
	\lambda_1 >0,\ \lambda_2>0,\ 4\lambda_1\lambda_2 - \lambda_3^2\, ,\label{eq:stabcondition}
\end{equation}
which we will investigate in the next section.

\section{Effect of kinetic mixing at two-loop\label{sec:running}}

The RGEs for the SM B-L as derived with \pyrate have been given recently in~\cite{Lyonnet:2016xiz} which we refer to for the full expressions and details. Our goal here is not to obtain precise conclusions regarding the physics of the model at hand. It is rather to quantify the impact of kinetic mixing one can expect in a physical situation on the running of the parameters. Therefore, we neglect the electroweak matching conditions as well as the scalar threshold corrections due to the heavy Higgs. These corrections would not affect our conclusions regarding the amplitude of the impact of the kinetic mixing. The initial values of all the parameters are set at the scale of the Z-boson mass, $M_{Z}$, and evolved to $10^{19}$ GeV using \CC routines as provided by the \pyrate package.

\subsection{Running of the parameters}

\begin{figure*}[!ht]
	\centering
	\includegraphics[scale=0.4]{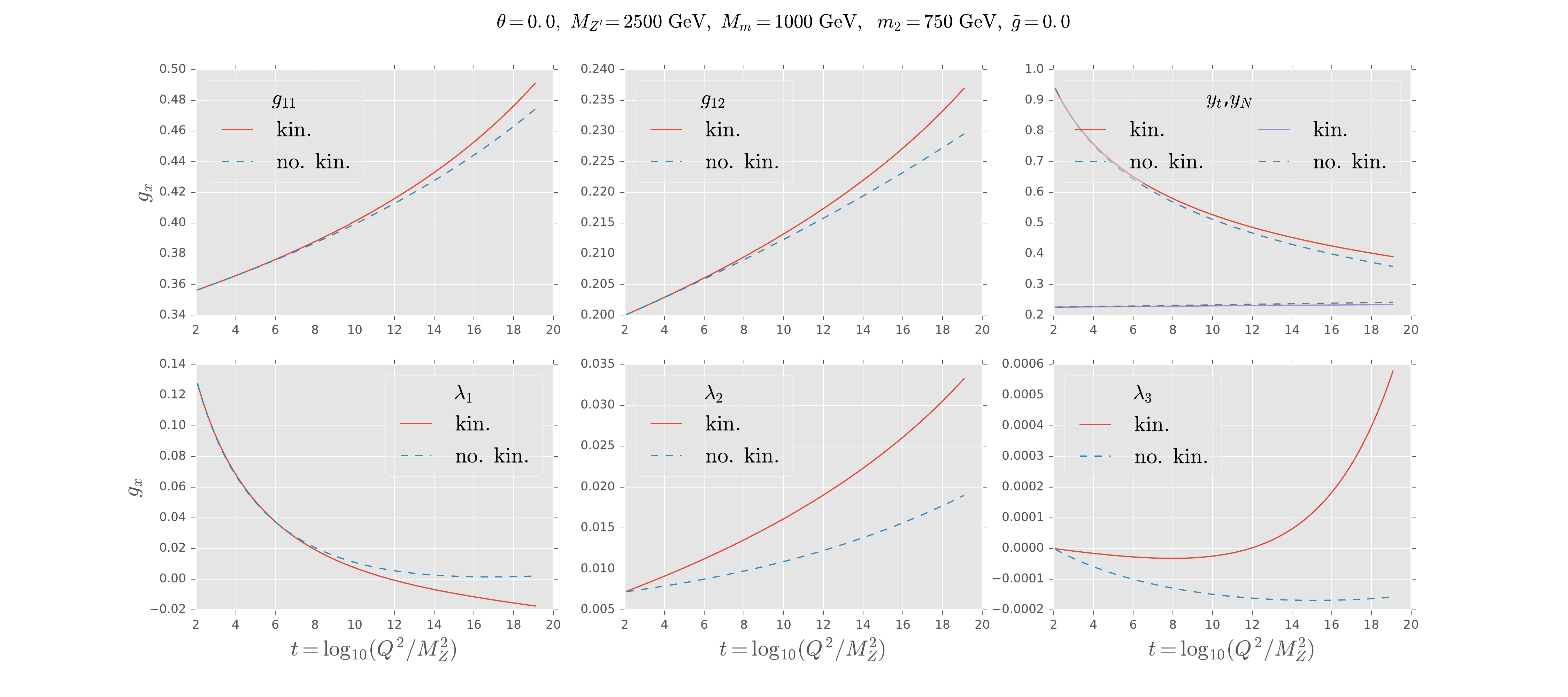}
	\caption{Running of the parameters in the B-L model in the case where the kinetic mixing is taken into account (solid lines) or neglected (dashed lines), for the benchmark point $\mathcal{B}_1$. See text and figure title for the value of the input parameters. The parameters plotted are $g_{1},g_{12},y_{t},y_{N},\lambda_{1},\lambda_{2},\lambda_{3}$.}
	\label{fig:kinmix1}
\end{figure*}

To begin, we simplify the situation and select a scenario in which there is no mixing between the two scalars, i.e. $\theta=0$. In addition, we set $\tilde{g}=0$ and will investigate the effect of $\tilde{g}\neq 0$ later. Note that this does not correspond to neglecting kinetic mixing as it is generated automatically by radiative corrections. The mass of the new heavy gauge boson $M_{Z'}$ is set to $2.5$~TeV which is in the ballpark of the recent exclusion limits derived in~\cite{Klasen:2016qux}. Finally, we set the mass of the heavy Higgs to 750 GeV. Note that the value of the input parameters are chosen to be representative of the situation but are not fine tuned in any way and the size of the effects obtained with different numerical values are very similar. This first benchmark point will be referred to as $\mathcal{B}_1=\{0, 2500, 1000, 0.2, 0, 126, 750\}$, see Eq.(\ref{eq:benchdef}) where the dimensionful parameters are given in GeV.

\begin{figure*}[!ht]
	\centering
	\includegraphics[scale=0.4]{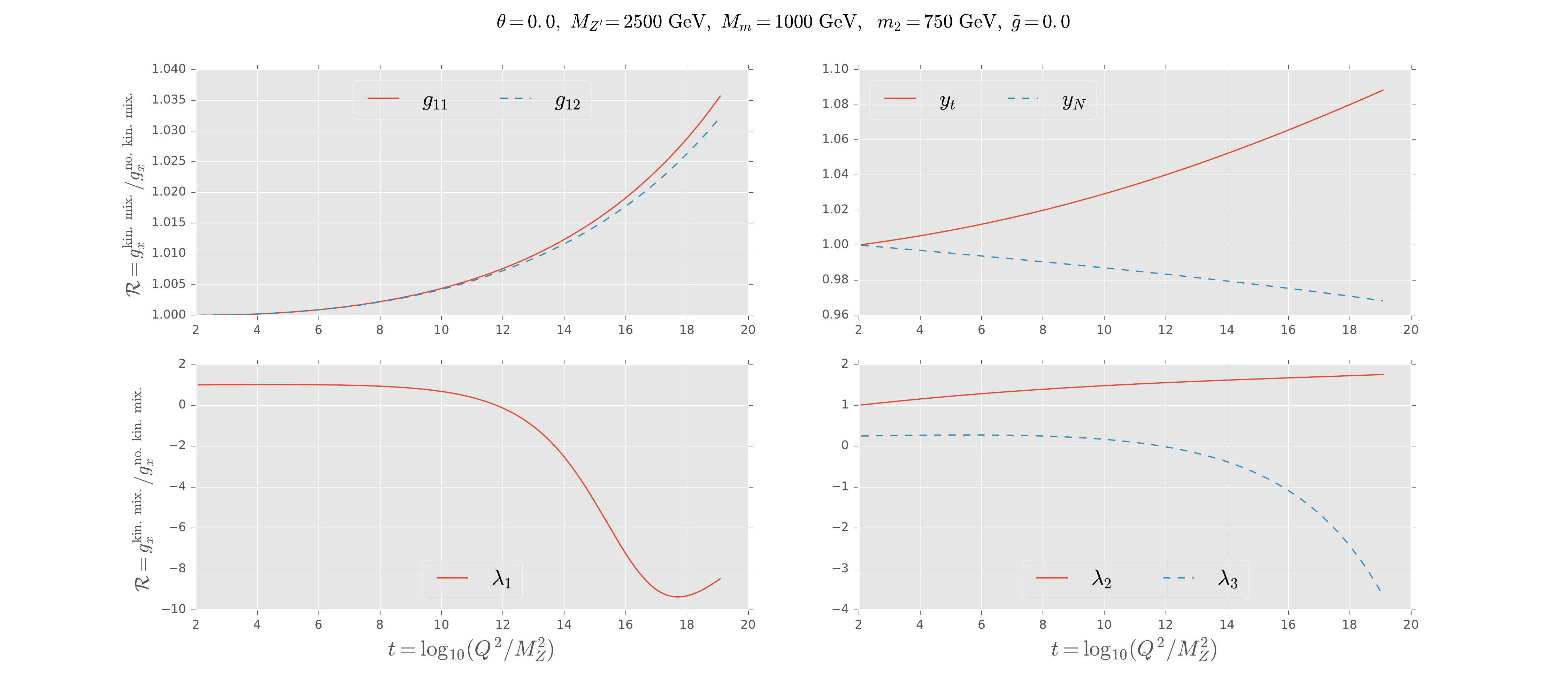}
	\caption{Ratio of the couplings taking into account or neglecting the kinetic mixing. The parameters plotted are $g_{1},g_{12},y_{t},y_{N},\lambda_{1},\lambda_{2},\lambda_{3}$. The initial values are those of $\mathcal{B}_1$.}
	\label{fig:kinmix1-ratio}
\end{figure*}

Fig.~\ref{fig:kinmix1} shows the running of some of the parameters of the B-L model when the kinetic mixing is fully taken into account at two-loop or neglected. More specifically, the gauge couplings $g_{11}$ and $g_{12}$, the quartic couplings $\lambda_1,\ \lambda_2,\ \lambda_3$ and the Yukawa couplings $y_t$ and $y_N$ are shown. On Fig.~\ref{fig:kinmix1-ratio} we display the corresponding ratio of the beta functions including kinetic mixing over the beta functions neglecting kinetic mixing.

\begin{figure}[!ht]
	\centering
	\includegraphics[width=0.45\textwidth]{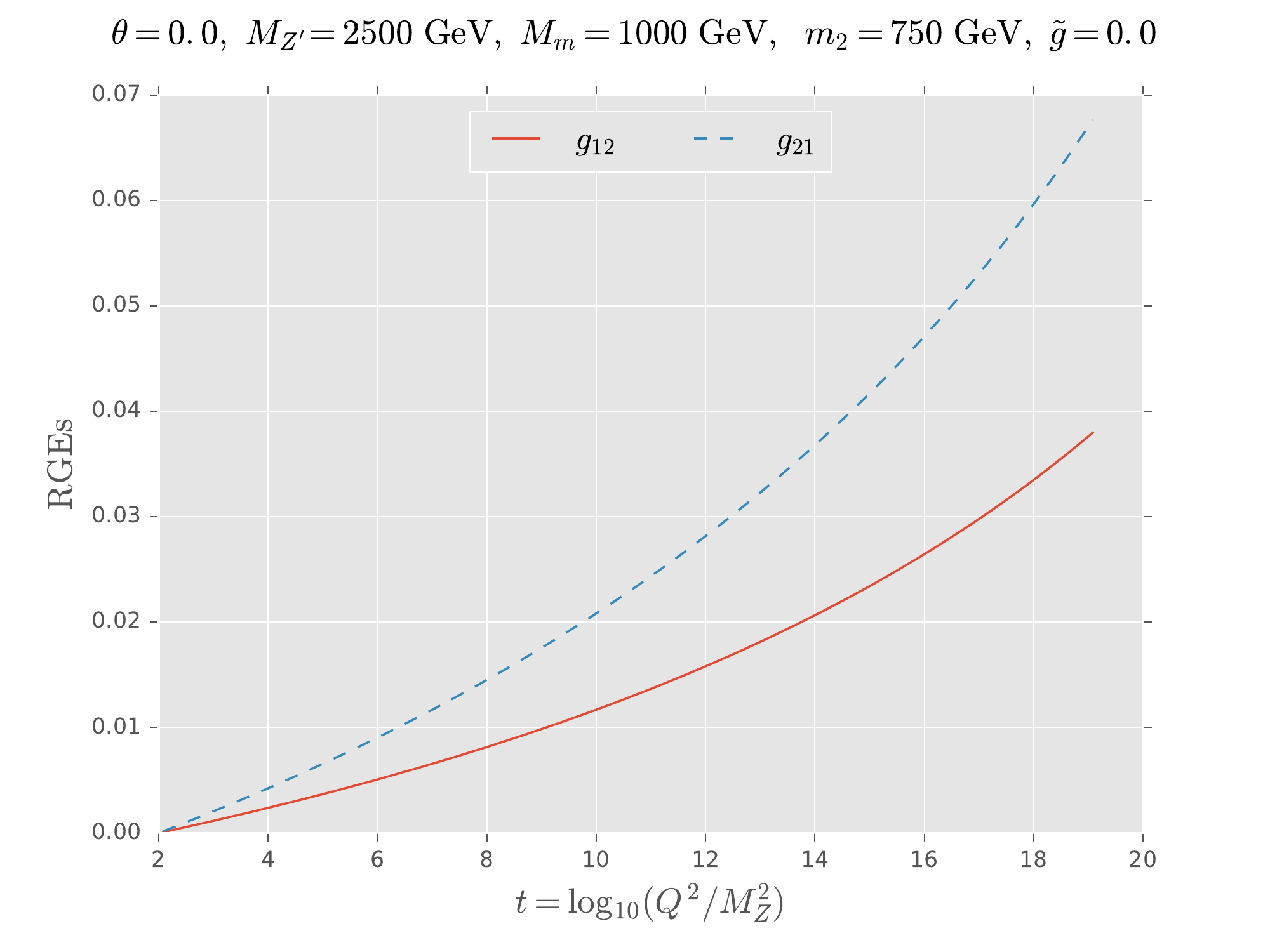}
	\caption{Running of the Abelian off-diagonal gauge couplings, $g_{12}$ and $g_{21}$. The input parameters are the same as in Fig.~\ref{fig:kinmix1}.}
	\label{fig:gaugerunning}
\end{figure}

While the impact on the gauge couplings is somewhat limited to 1-3 \% it reaches 2-6\% for the Yukawa couplings at the GUT scale. For the quartic couplings, the change is dramatic and deserves some comments.
\begin{enumerate}[(i)]
	\item $\lambda_1$ actually goes to zero when the kinetic mixing is neglected, see~Fig.~\ref{fig:kinmix1}, causing the ratio to blow up;
	\item $\lambda_{2}$ gets large kinetic contributions at one-loop of the form $\sim g^{2}\lambda$ and $\sim g^{4}$ coming from $\Lambda^S_{abcd}$ and $A_{abcd}$ in the notation of~\cite{Machacek:1984zw}. Moreover, there is a back reaction coming from the impact of kinetic mixing on $\lambda_3$ since $\beta_{\lambda_2}\sim 2\lambda_3^{2}$. These effects ultimately lead to a ratio of 1.5 at $10^{10}$ GeV; 
	\item the kinetic mixing in $\lambda_3$ is such that it turns the sign of the beta function around, from negative to positive at a scale of about $10^{8}$ GeV, ultimately leading to a positive $\lambda_3$ at a scale of $10^{12}$ GeV.
\end{enumerate}
Therefore, as expected the impact of kinetic mixing is governed by the evolution of the off-diagonal effective Abelian gauge couplings, $g_{12},\ g_{21}$. In Fig.~\ref{fig:gaugerunning}, we show the running of these two gauge couplings with the energy. It is important to note that the values of these gauge couplings stay perturbative all the way up to the Planck scale and that the effects seen in the other parameters are not the result of extreme values of the gauge couplings.

\subsection{Stability of the potential}

As we have already seen, the impact of kinetic mixing can be quite large on the running of the different parameters. In this section, we investigate how this translates in terms of the stability conditions, Eq.~(\ref{eq:stabcondition}). 

\begin{figure}[!ht]
	\centering
	\includegraphics[width=0.45\textwidth]{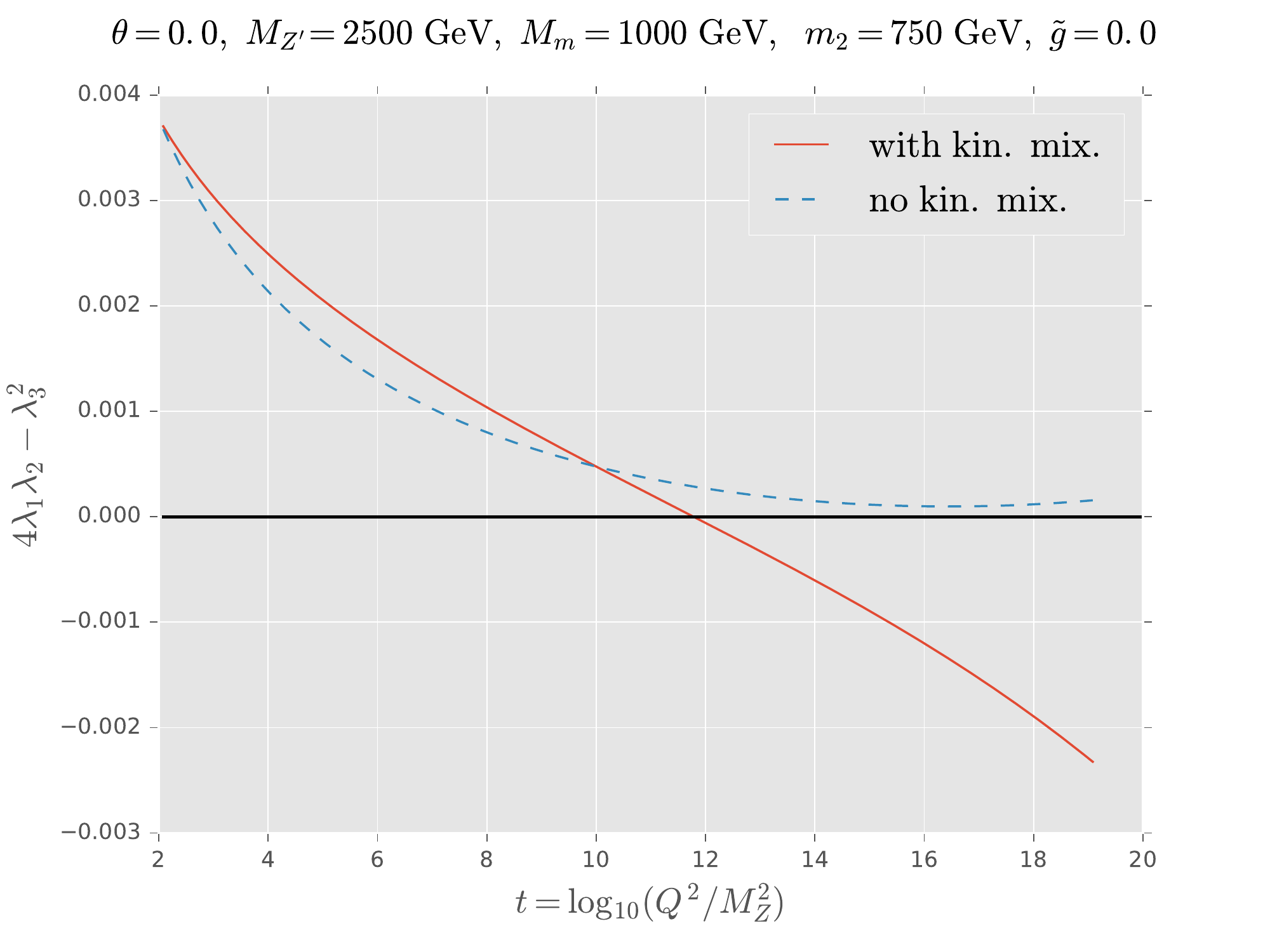}
	\caption{Stability condition for the benchmark point $\mathcal{B}_1$ of the B-L model in the case where the kinetic mixing is (solid line) or is not (dashed line) taken into account. The impact of kinetic mixing in this case leads to different conclusions.}
	\label{fig:stability1}
\end{figure}

Fig.~\ref{fig:stability1}, shows the stability condition $4\lambda_1\lambda_2-\lambda_3^2$ with and without kinetic mixing for $\mathcal{B}_1$. The difference in this case is striking as the scenario develops an instability at around $10^{12}$ GeV when kinetic mixing is accounted for.
This behaviour is actually simple to understand as the instability is the direct consequence of $\lambda_1$ turning negative at around the same scale, $\Lambda\sim 10^{12}$ GeV.

\begin{figure}[!ht]
	\centering
	\includegraphics[width=0.45\textwidth]{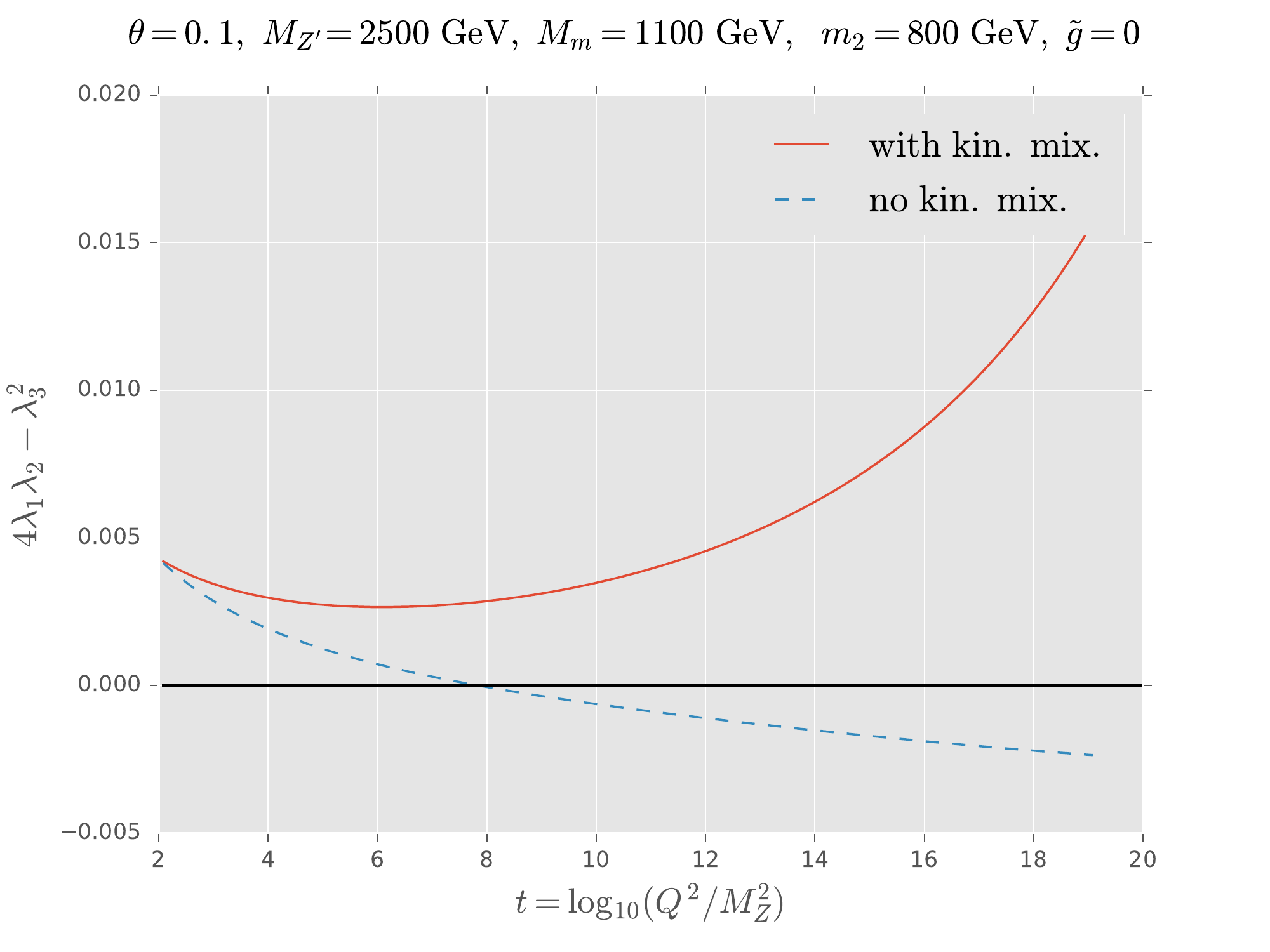}
	\caption{Stability condition for $\mathcal{B}_2$. In this example, the kinetic mixing {\em rescues} the stability of the potential}
	\label{fig:stability2}
\end{figure}

One might think that such a drastic change in results is limited to a small region of the parameter space, however that is not the case. Indeed, one can easily find regions where even the opposite situation happens, i.e. where the kinetic mixing {\em rescues} the stability of the potential. Fig.~\ref{fig:stability2} shows such an example, for $\mathcal{B}_2=\{0.1,2500,1100,0.2,0,126,800\}$. This scenario is the result of several competing effects.
\begin{enumerate}[(i)]
	\item $\theta\neq 0$ leads to $\lambda_3^{\mathcal{B}_2}(M_{Z})\gg\lambda_3^{\mathcal{B}_1}(M_Z)$ which remains true at all scales, see Eq.~\ref{eq:lbds}.
	\item $\lambda_{1}(M_{Z})\sim m_{2}^2/(4v^2)(1-\cos2\theta)$, therefore $\theta\neq 0$ greatly increases $\lambda_1(M_{Z})$ which leads to $\lambda_1>0$ at all scales.
	\item Finally, $\lambda_2$ gets large kinetic corrections sufficient to overcome the increase in $\lambda_3$.
\end{enumerate}

One more thing to note is that the mass of the heavy Higgs plays a crucial role here. The initial values of $\lambda_{1,2,3}$ decrease with $m_{2}^{2}(M_{Z})$ and for $m_{2}^{2}(M_{Z})\sim500$ GeV or lighter, $\lambda_2$ quickly runs negative without kinetic mixing leading to an unstable potential.

\subsection{Impact of $\tilde{g}$} 

\begin{figure}[!ht]
	\centering
	\includegraphics[scale=0.4]{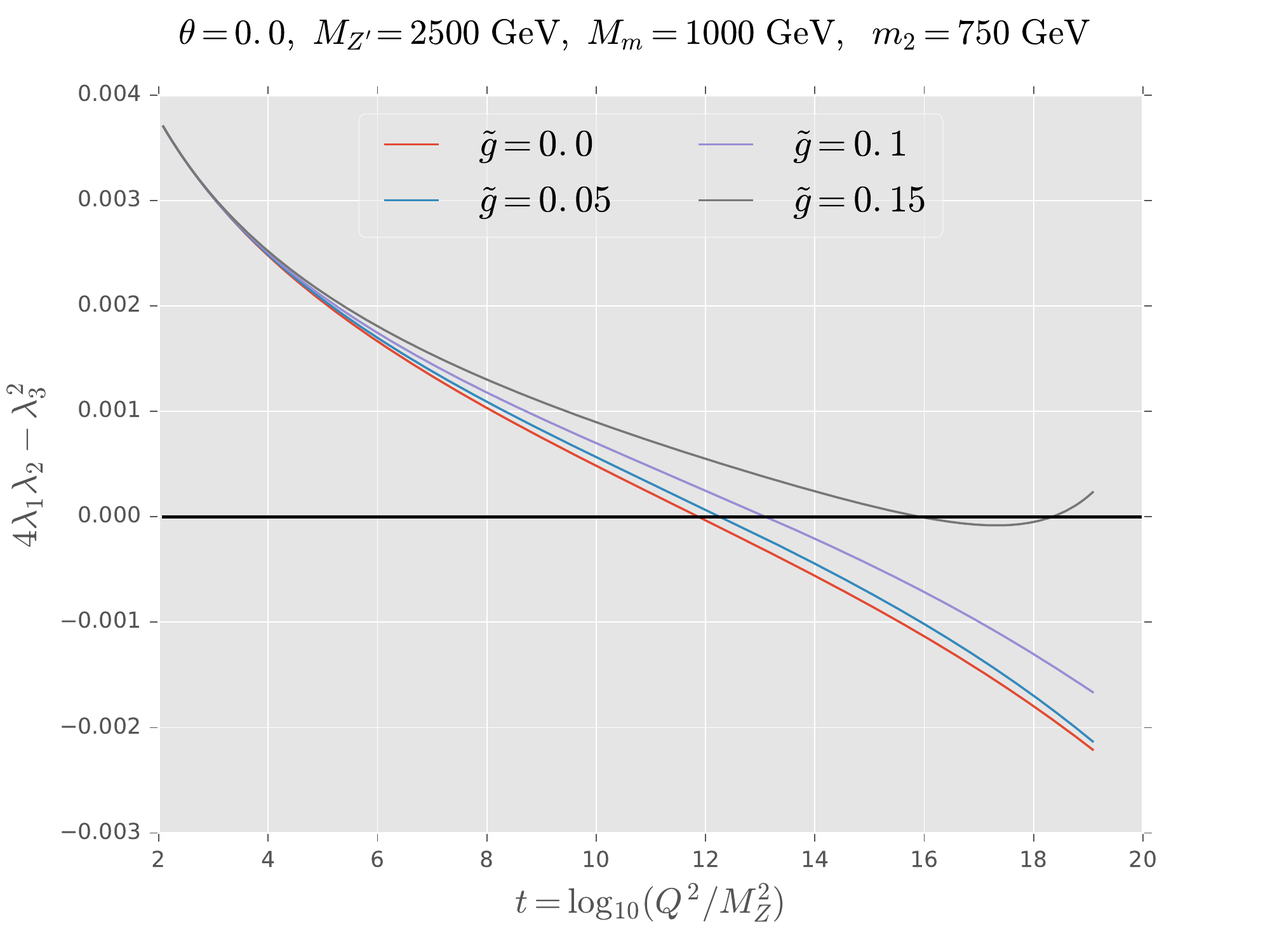}
	\caption{Stability condition in the $\mathcal{B}_1$ benchmark point for several values of $\tilde{g}\in\{0,0.05,0.1,0.15\}$.}
	\label{fig:kinmixstabgt}
\end{figure}

\begin{figure*}[!ht]
	\centering
	\includegraphics[scale=0.4]{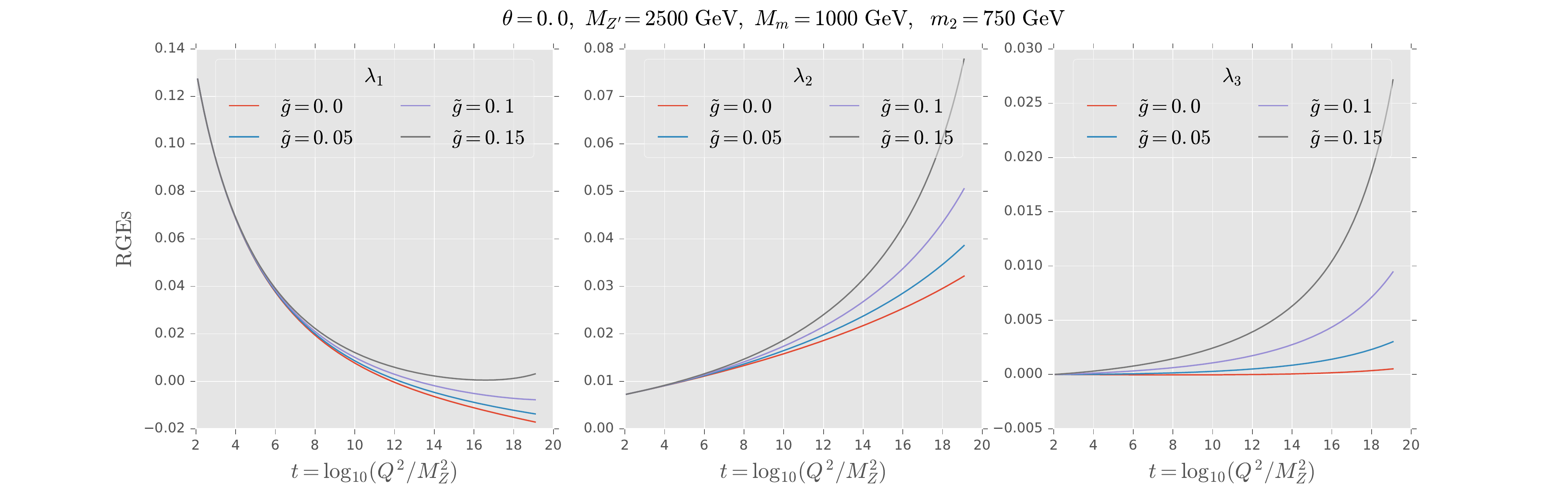}
	\caption{Running of the quartic couplings in the $\mathcal{B}_1$ benchmark point for several values of $\tilde{g}\in\{0,0.05,0.1,0.15\}$.}
	\label{fig:kinmixgt}
\end{figure*}

Finally, we investigate the impact of the value of $\tilde{g}$ on the running of the parameters and the stability of the potential. To do so, we consider $\mathcal{B}_1$ but with $\tilde{g}\in\{0.,0.05,0.1,0.15\}$. The results are presented in Figs.~\ref{fig:kinmixstabgt} and \ref{fig:kinmixgt} in which we show the running of the stability condition and the corresponding quartic couplings for different values of $\tilde{g}$ respectively. Values of $\tilde{g}>0.15$ will lead to non-perturbative couplings at the highest energies around $10^{19}$ GeV (not shown). The spread in $\lambda_{1,2}$ values, Fig.~\ref{fig:kinmixgt}, due to different values of $\tilde{g}$ is 35\% and 15\% respectively at the scale $10^{10}$ GeV while the $\lambda_{3}$ parameter is extremely small for $\tilde{g}=0$ resulting in a spread of two orders of magnitude at the same scale.

\subsection{Two-loop kinetic mixing contribution}

\begin{figure}[!ht]
	\centering
	\includegraphics[width=0.45\textwidth]{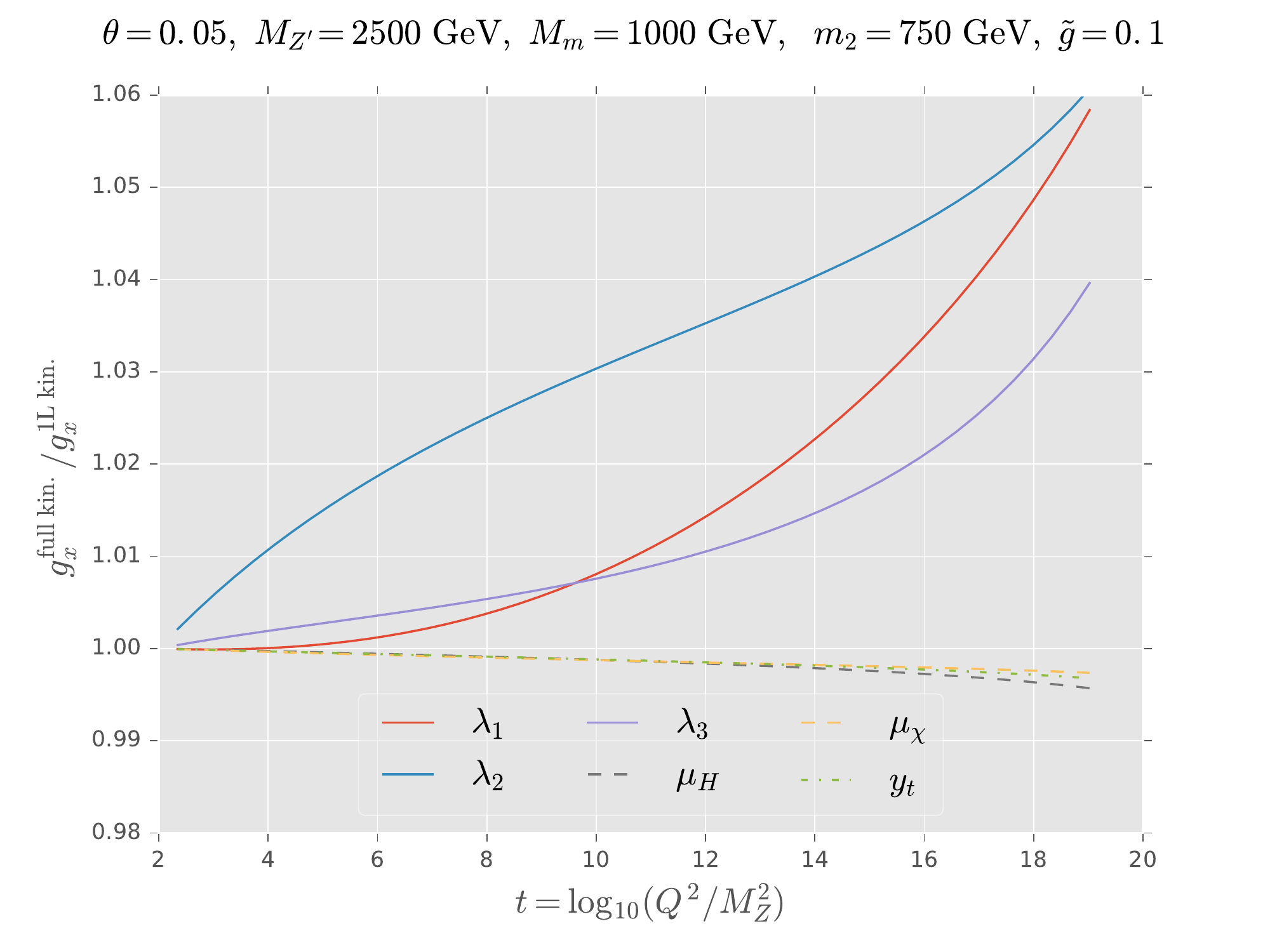}
	\caption{Running of some of the couplings of the B-L model for benchmark point $\mathcal{B}_3$. We show the ratio of predictions including the full two-loop corrections coming from kinetic mixing over those including only the corrections up to two-loop order in the gauge couplings and one-loop elsewhere.}
	\label{fig:xikin}
\end{figure}

\begin{figure}[!ht]
	\centering
	\includegraphics[width=0.45\textwidth]{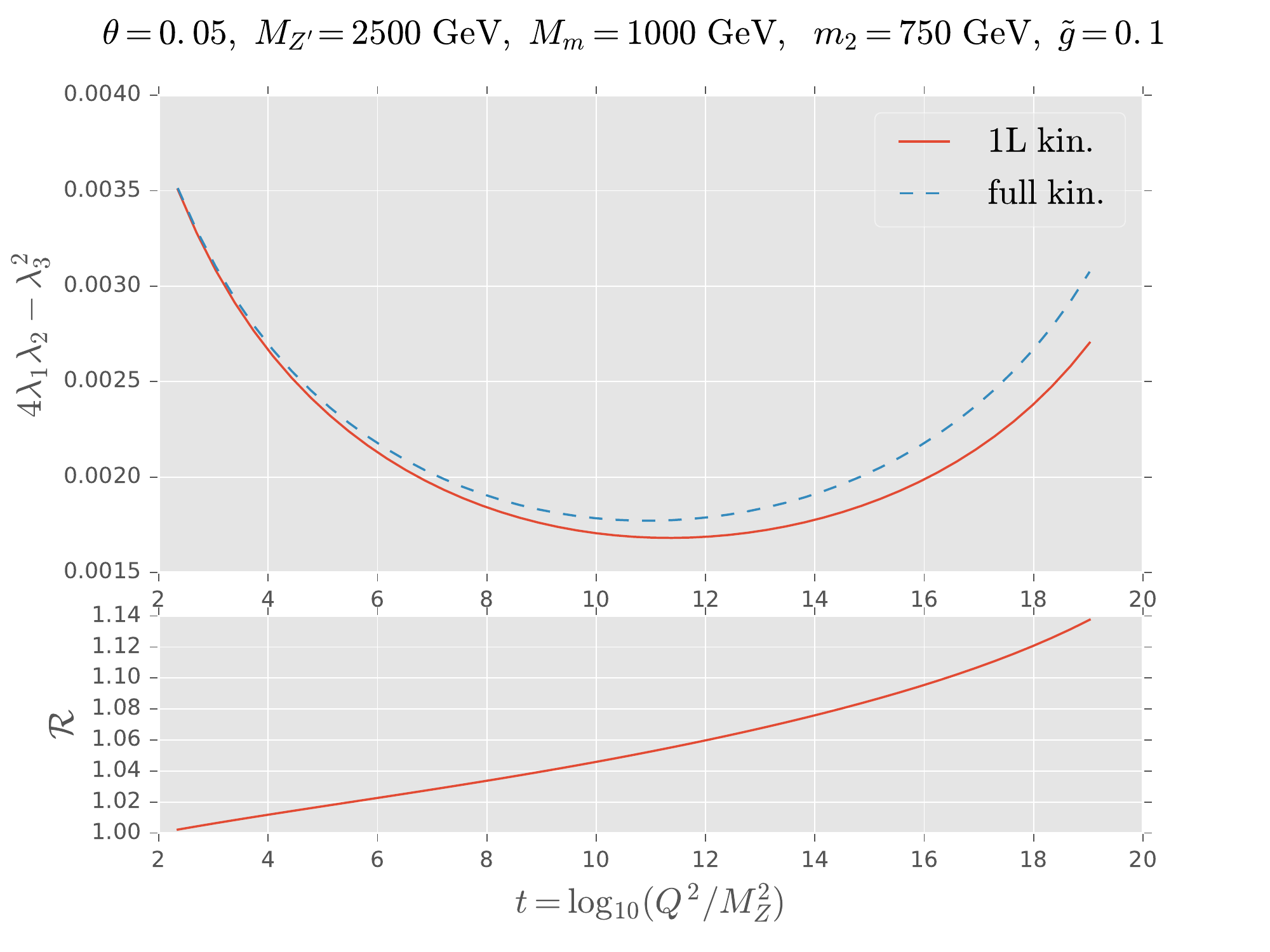}
	\caption{Stability condition in the B-L model for benchmark point $\mathcal{B}_3$. We  show the predictions including the full two-loop corrections (dashed line) coming from kinetic mixing and those including only the corrections up to two-loop order in the gauge couplings and one-loop elsewhere (solid line).}
	\label{fig:xikinstab}
\end{figure}

In this last part we focus on determining the amplitude of the kinetic mixing contributions at two-loop versus the one-loop order ones. To do so, we consider the two-loop beta functions without kinetic mixing to which we add either the two-loop, or one-loop order corrections coming from kinetic mixing. For simplicity, we keep the kinetic mixing contribution up to two-loop in the gauge couplings but switch on and off the two-loop order contributions in the other parameters. This would correspond to a simplified implementation of the rules of~\cite{Fonseca:2013jra} in which the more involved replacement rules are ignored, and in particular the ones involving the scalar generators. 

Fig.~\ref{fig:xikin} shows ratios of couplings at two-loop as a function of the scale, in which the numerator is the coupling including the whole set of two-loop kinetic contributions, while the denominator is the same coupling in which only the one-loop kinetic mixing contributions have been retained~\footnote{Note that for the gauge couplings, the two-loop contributions are also taken into account.}. The benchmark point, $\mathcal{B}_3$, used here is defined by $\mathcal{B}_3=\{0.05,2500,750,0.2,0.1,126,1000\}$. The ratios are below 1\% for the parameters $\mu_H,\mu_\chi$ and $y_N$, whereas for the quartic terms, they are typically larger than 1\% at the scale $10^{10}$ GeV and can reach 5\% at the GUT scale. 

The corresponding effect on the stability condition, Eq.(\ref{eq:stabcondition}), are shown in Fig.~\ref{fig:xikinstab}. In the upper panel, the dashed line represent the stability condition in the case where all the corrections are included while the solid line is the result of retaining only the one-loop kinetic contribution. From the bottom panel, showing the ratio, it is easy to extract that the difference in the two approaches can be of the order 10\%. 

While $\mathcal{B}_3$ is a particular point in the parameter space, it illustrates that neglecting the two-loop contributions coming from kinetic mixing might lead to an error of the order a couple of percents. In addition, it is not unlikely, that in some combinations of couplings like the stability condition of Eq.(\ref{eq:stabcondition}) these differences are enhanced leading to larger discrepancies. Furthermore, since contributions from the kinetic mixing terms and the base beta functions mix together, a coherent perturbative treatment at two-loop demands to take into account contributions coming from kinetic mixing at the same order.

\section{Conclusion\label{sec:conclusion}}

Kinetic mixing is a fundamental property of models with an extended Abelian gauge structure like the SM B-L. Taking into account the kinetic mixing in the RGEs at two-loop is a complex procedure. To this end, we have consistently implemented the kinetic mixing at two-loop in the software \pyrate. Therefore, these models now benefit from the same level of automation as their counterparts without kinetic mixing. 

After reviewing the theoretical setup of the SM B-L, we studied in detail the impact of kinetic mixing on the running of the parameters of the model. We showed that neglecting the kinetic mixing can lead to erroneous conclusions regarding the stability of the scalar potential. In addition, we studied the impact of the kinetic mixing contributions at two-loop and showed that they can be significant with respect to their one-loop counterparts. 

Of course, our goal was only to illustrate that the features of a model can dramatically change when properly accounting for kinetic mixing; a careful treatment of the electroweak matching conditions and scalar threshold corrections would be required to draw detailed physical conclusions regarding the B-L model investigated here. We leave this to future work. 

We believe that the order of magnitude of the effects exemplified with the SM B-L can be similar in other models and that taking kinetic mixing into account is crucial to obtain meaningful results. This is now a simple task thanks to \pyrate. 

\section*{Acknowledgments}
I would like to thank T. Jezo,  A. Kusina, F. Olness, I. Schienbein, and F. Staub for their useful comments and for reviewing this manuscript. This work was also partially supported by the U.S. Department of Energy under Grant No. DE-SC0010129.

\bibliography{mybibliography}
\bibliographystyle{utphys}

\end{document}